\documentclass[12pt,preprint]{aastex}
\usepackage{natbib}
\usepackage{graphicx}
\usepackage{latexsym}
\usepackage{epsfig} 
\usepackage{amssymb,amsbsy}
\usepackage{amsmath}
\usepackage{eucal}
\usepackage[dvipsnames]{color}

\shorttitle{Primordial Planets}
\shortauthors{Shchekinov, Safonova \&  Murthy}

\begin{document}

\newcommand{\3}{\ss}
\newcommand{\n}{\noindent}
\newcommand{\eps}{\varepsilon}
\newcommand{\be}{\begin{equation}}
\newcommand{\ee}{\end{equation}}
\def\ba{\begin{eqnarray}}
\def\ea{\end{eqnarray}}
\def\de{\partial}
\def\msun{M_\odot}
\def\div{\nabla\cdot}
\def\grad{\nabla}
\def\rot{\nabla\times}
\def\ltsima{$\; \buildrel < \over \sim \;$}
\def\simlt{\lower.5ex\hbox{\ltsima}}
\def\gtsima{$\; \buildrel > \over \sim \;$}
\def\simgt{\lower.5ex\hbox{\gtsima}}
\def\fr{\frac}
\def\la{\langle}
\def\ra{\rangle}
\def\var{\varphi}
\def\half{\frac{1}{2}}
\def\quart{\frac{1}{4}}
\def\third{\frac{1}{3}}
\def\quart{\frac{1}{4}}
\def\fifth{\frac{1}{5}}
\def\sixth{\frac{1}{6}}
\def\red{\textcolor{red}}
\def\lhs{left-hand side\,\,}
\newcommand{\rhs}{right-hand side\,\,}
\newcommand{\ns}{\!\!}
\newcommand{\tc}{\textcircled}
\newcommand{\ddel}[2]{\frac{\partial{#1}}{\partial{#2}}}
\newcommand{\dd}[1]{\frac{\partial}{\partial {#1}}}
\newcommand{\dder}[2]{\frac{d{#1}}{d{#2}}}
\newcommand{\tens}[1]{\buildrel {\textstyle \rightrightarrows}\over {#1}}
\newcommand{\ontop}[1]{\buildrel {(n)}\over {#1}}
\newcommand{\llsim}{\genfrac{}{}{0pt}{2}{\ll}{\sim}}
\newcommand{\pol}[1]{\stackrel{\rm LCP}{\mathrm{RCP}}}
\newcommand{\tx}{\ensuremath{\underset{\textstyle\small\sim}{x}}}
\newcommand{\rss}{\scriptscriptstyle}
\newcommand{\app}{\approx}
\newcommand{\del}{\partial}
\newcommand{\bit}[1]{\ensuremath{\textit{\bfseries{#1}}}}
\newcommand{\bs}{\boldsymbol}
\newcommand{\me}{\mathrm{e}}
\newcommand{\mi}{\mathrm{i}}
\newcommand{\no}{\nonumber}
\newcommand{\und}{\underline}
\renewcommand{\a}{\alpha}
\renewcommand{\b}{\beta}
\newcommand{\g}{\gamma}
\newcommand{\G}{\Gamma}
\renewcommand{\d}{\delta}
\newcommand{\D}{\Delta}
\renewcommand{\l}{\lambda}
\renewcommand{\L}{\Lambda}
\renewcommand{\o}{\omega}
\renewcommand{\O}{\Omega}
\renewcommand{\r}{\rho}
\newcommand{\s}{\sigma}
\newcommand{\Sig}{\Sigma}
\renewcommand{\t}{\theta}
\renewcommand{\k}{\kappa}
\newcommand{\om}{o}

\title{Planets in the Early Universe}

\author{Yu.~A.~Shchekinov}
\affil{Department of Space Physics, SFU, Rostov on Don, Russia}
\email{yus@sfedu.ru}

\author{M.~Safonova,~J.~Murthy}
\affil{Indian Institute of Astrophysics, Bangalore, India}
\email{rita@iiap.res.in, 
murthy@iiap.res.in}

\begin{abstract}

Several planets have recently been discovered around stars that are
old and metal-poor, implying that these planets are also old, formed 
in the early Universe together with their hosts. The canonical theory 
suggests that the conditions for their
formation could not have existed at such early epochs. In this paper
we argue that the required conditions, such as sufficiently high
dust-to-gas ratio, could in fact have existed in the early Universe 
immediately following the first episode of metal production in Pop.~III
stars, both in metal-enhanced and metal-deficient environments. 
Metal-rich regions may have existed in multiple isolated pockets
of enriched and weakly-mixed gas close to  the massive Pop.~III stars.
Observations of quasars at redshifts $z\sim 5$, and gamma-ray bursts at
$z\sim 6$, show a very wide spread of metals in absorption from $\rm [X/H]
\simeq -3$ to $\simeq -0.5$. This suggests that physical conditions in
the metal-abundant clumps could have been similar to where protoplanets
form today. However, planets could have formed even in low-metallicity
environments, where formation of stars is expected to proceed due to lower
opacity at higher densities. In such cases, the circumstellar accretion
disks are expected to rotate faster than their high-metallicity analogues.
This in turn can result in the enhancement of dust particles at the disk
periphery, where they can coagulate and start forming planetesimals. In
conditions with the low initial specific angular momentum of the cloud,
radiation from the central protostar can act as a trigger to drive
small-scale instabilities with typical masses in the Earth to Jupiter
mass range. Discoveries of planets around old metal-poor stars (e.g. HIP
11952, $\rm [Fe/H]\sim -1.95$, $\sim 13$ Gyr) show that planets did indeed form in
the early Universe and this may require modification of our understanding
of the physical processes that produce them. This work is an attempt to
provide one such heuristic scenario for the physical basis for their
existence. 
\end{abstract}


\section{Introduction}
\label{sec:1}

\noindent 

A two-year gravitational lensing survey by the Microlensing Observations in Astrophysics (MOA) and 
Optical Gravitational Lensing Experiment (OGLE) groups towards the Galactic bulge (Sumi et al. 2011) 
has found ten events which can be attributed to Jupiter-mass planets. These microlensing planets are 
free-floating, in the sense that no host stars have been detected within about 10 AU. Recent 
estimates show that there can be up to $10^5$ as many such planets as stars in the Galaxy 
\citep{s11,nomads2012}; a more conservative estimate has been found by Tutukov \& Fedorova (2012). 
Three out of the 10 microlensing events from planets have galactic latitudes of $b\simeq -3^\circ$, 
and one has $b\simeq -6^\circ$, corresponding to the heights from 0.4 to 0.8 kpc. Therefore these 
planets are likely to belong to the thick disk population, where a considerable fraction 
(up to 30\% to 50\%) of stars is occupied by the oldest 
(Pop.~II) Milky Way stars with mean metallicity\footnote{$\rm [X/H]=
\log_{10}{(N_{\rm X}/N_{\rm H}})_{star}-\log_{10}{(N_{\rm X}/N_{\rm H}})_{\sun}$.}
$\langle[{\rm Fe/H}]\rangle=-0.6$ \citep{z08,b11,h11}. In spite of a generally uncertain 
age-metallicity relationship (see, e.g. Soderblom 2010), 
F--G dwarf thick disk stars with $\langle[{\rm Fe/H}]\rangle=-0.6$ are found to have ages 
with a relatively narrow spread, $10\pm$2 Gyr (Feltzing \& Bensby 2008), although 
more recent studies (see, e.g. Casagrande et al. 2011) advocate a wider spread, 5 to 13 Gyr
at $\langle[{\rm Fe/H}]\rangle=-0.6$, which however quickly converges to $11\pm 2$ Gyr at 
$\langle[{\rm Fe/H}]\rangle\leq-0.7$.Therefore these free-floating
planets might also be old, formed in the early Milky Way Galaxy more than 10 Gyr ago;  
their free-floating status is a result of being ejected from the parent 
systems by planet-planet scattering events during the early stages of planetary systems life 
(e.g., Ford, Rasio \& Yu 2006, Tutukov \& Fedorova 2012). 

\subsection{The ``Metallicity Effect''}

In general, a well-known planet distribution function peaking at solar metallicity -- 
the ``metallicity effect'' (Udry \& Santos 2007) -- is readily understood in terms 
of the thermodynamics of the gas from which the planets are formed: 
the smaller the gas metallicity, the weaker is the radiation cooling and therefore the higher are the 
gas temperature and the Jean's mass. In the extreme case of primordial gas, where the metallicity 
$\rm [Fe/H]=-\infty$, the only source of cooling is from molecular hydrogen lines, and only extremely
massive stars may be formed, $M > 80 \,M_{\sun}$ (eg. Schneider et al. 2004, Bromm \& Larson 2004), 
which would have evolved rapidly and be extinct by now. However, it was suggested (Nakamura \& Umemura 2001) 
that the fragmentation of filamentary primordial gas clouds could result in the formation of low-mass stars even in
the absence of metals, with only molecular hydrogen as the main cooling agent. Recent simulations (Clark et al. 2011,
Greif et al. 2011) demonstrated that the fragmentation process is rather environment-sensitive, 
such that gravitational instability may result in the formation of a group of low-mass ($M\sim 0.1-10~\msun$) 
protostars with a relatively flat initial mass function (IMF). The discovery of the extremely 
metal-poor ($\rm Z \leq 7.4\times 10^{-7}$)\footnote{Z is the mass fraction of metals} low-mass 
($M < 0.8~\msun$) 
star (Caffau et al. 2012) seems to confirm this conclusion. Formation of planetary fragments in 
gas with primordial chemical composition is nevertheless problematic, particularly because  H$_2$ and HD 
molecules provide quick thermalization of gas with the cosmic microwave background, which limits 
the fragment mass to $M\gtrsim 0.1~\msun$ \citep{sv06}.

\begin{figure}[h!]
\includegraphics[scale=0.3]{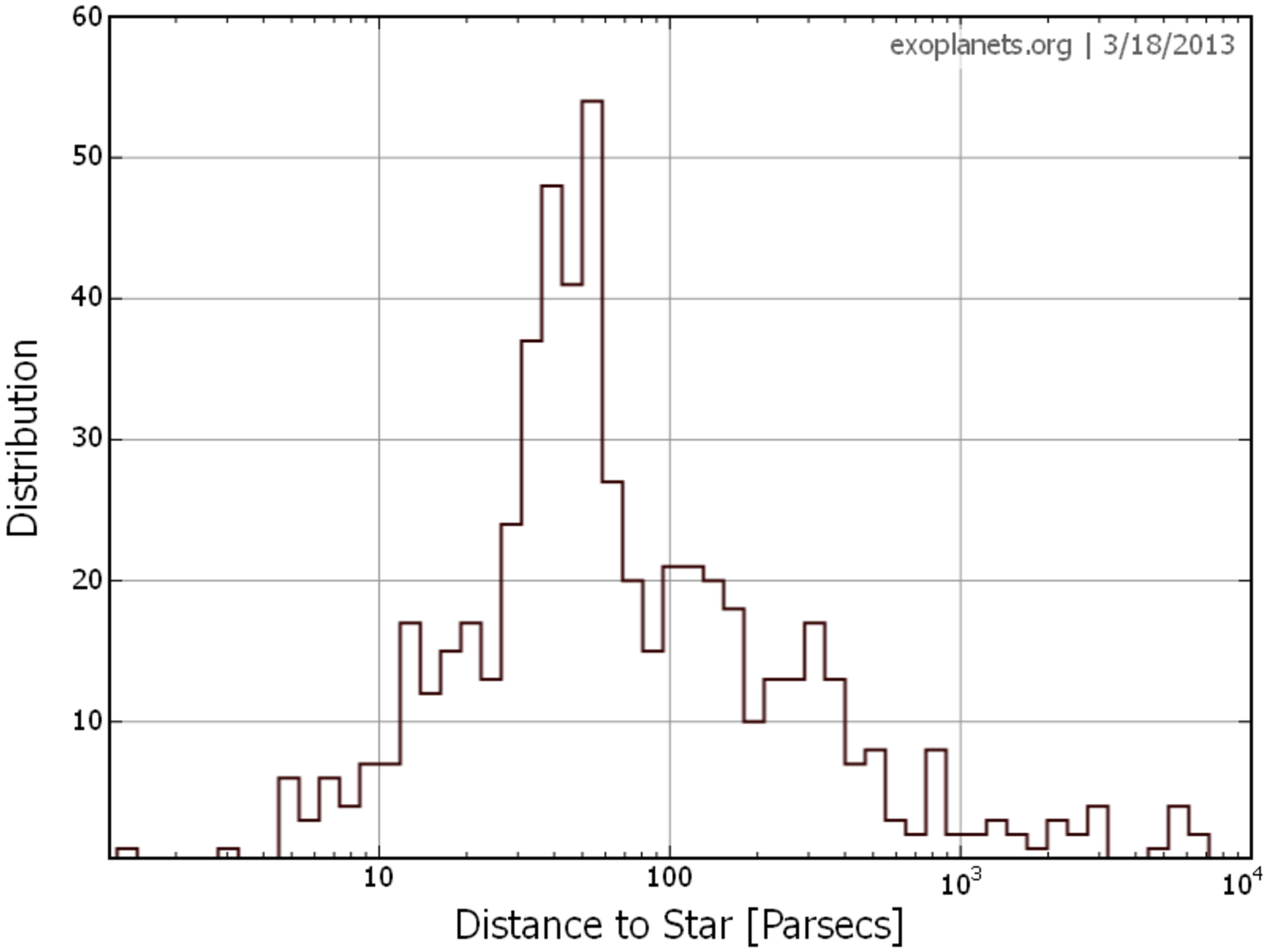}
\includegraphics[scale=0.65]{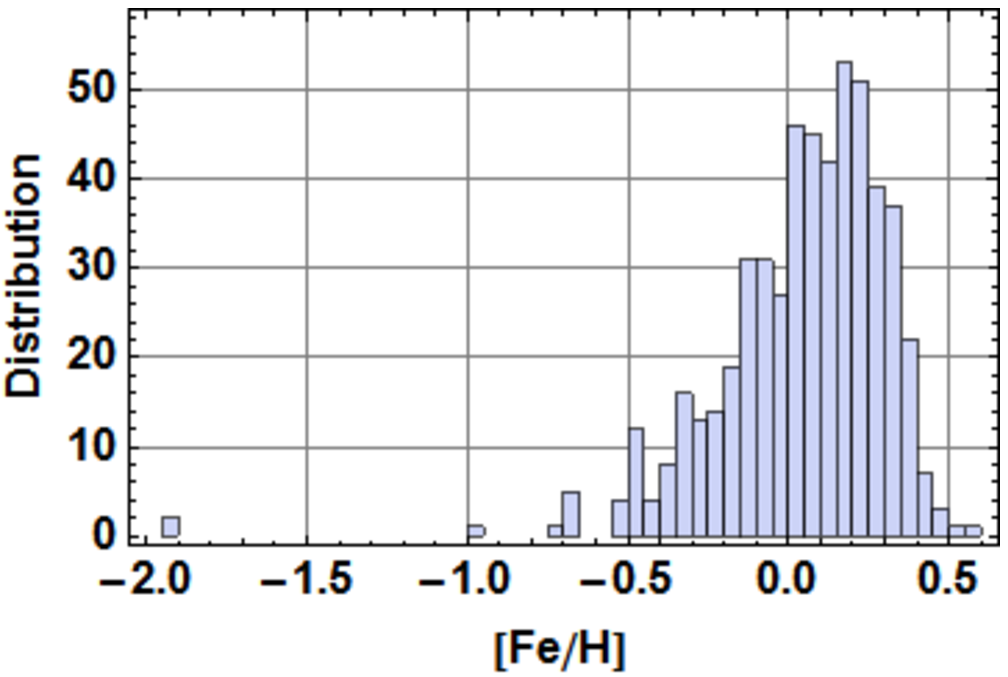}  
\caption{The left plot shows the distribution of stars with confirmed exoplanets (total 576 objects) 
with distance to the host. It is readily seen that the host distribution peaks at $\sim 60$ pc, 
and the majority are located in the solar vicinity, within 
$\lesssim 600$ pc of the Sun. On the right is the distribution of planets in the solar vicinity 
(within 600 parsec of the Sun) as function of host metallicity. Out of 535 planetary hosts with 
measured metallicity, 8 have $\rm [Fe/H] \leq -0.6$ ($\sim 1.5\%$). These plots indicate that
the observed ``metallicity effect'' may be a result of a strong observational selection effect.
This figure was made using the Exoplanet Orbit Database ({\it Left}) and the Extrasolar Planets 
Encyclopaedia data ({\it Right}).}
\label{fig:example}
\end{figure}

However, the observed ``metallicity effect''   may be a result of a strong observational selection effect 
in that metal-poor Pop.~II stars in the solar vicinity are at least two orders of magnitude less populous 
than those of solar metallicity (Gilmore et al, 1989; Dehnen \& Binney 1998, see Casagrande et al. 2011 
for more recent discussion). Classical detection methods (radial velocity measurements and transits) 
limit detection of planets to nearby stars (see Fig.~\ref{fig:example}, {\it Left}, for illustration) 
which predominantly belong to the thin disk. These stars have high metallicities and it is therefore a 
natural selection effect that planets are found around high-metallicity stars (Fig.~\ref{fig:example}, 
{\it Right}). This circumstance was first noted by Adibekyan 
et al. (2012a,b) from the abundances of $\alpha$-elements in planet-host stars. Moreover,
Adibekyan et al. (2012a) explicitly infer that metal-poor hosts belong mostly to the thick disk. 
Noticeably in this conection that all hosts beyond 600 pc have metallicities higher that 
$\rm [Z]=-0.4$, indicating that low-metallicity hosts are from a more extended sample 
where old stellar population is present. 

\begin{figure}[h!]
\begin{center}
\includegraphics[scale=0.36]{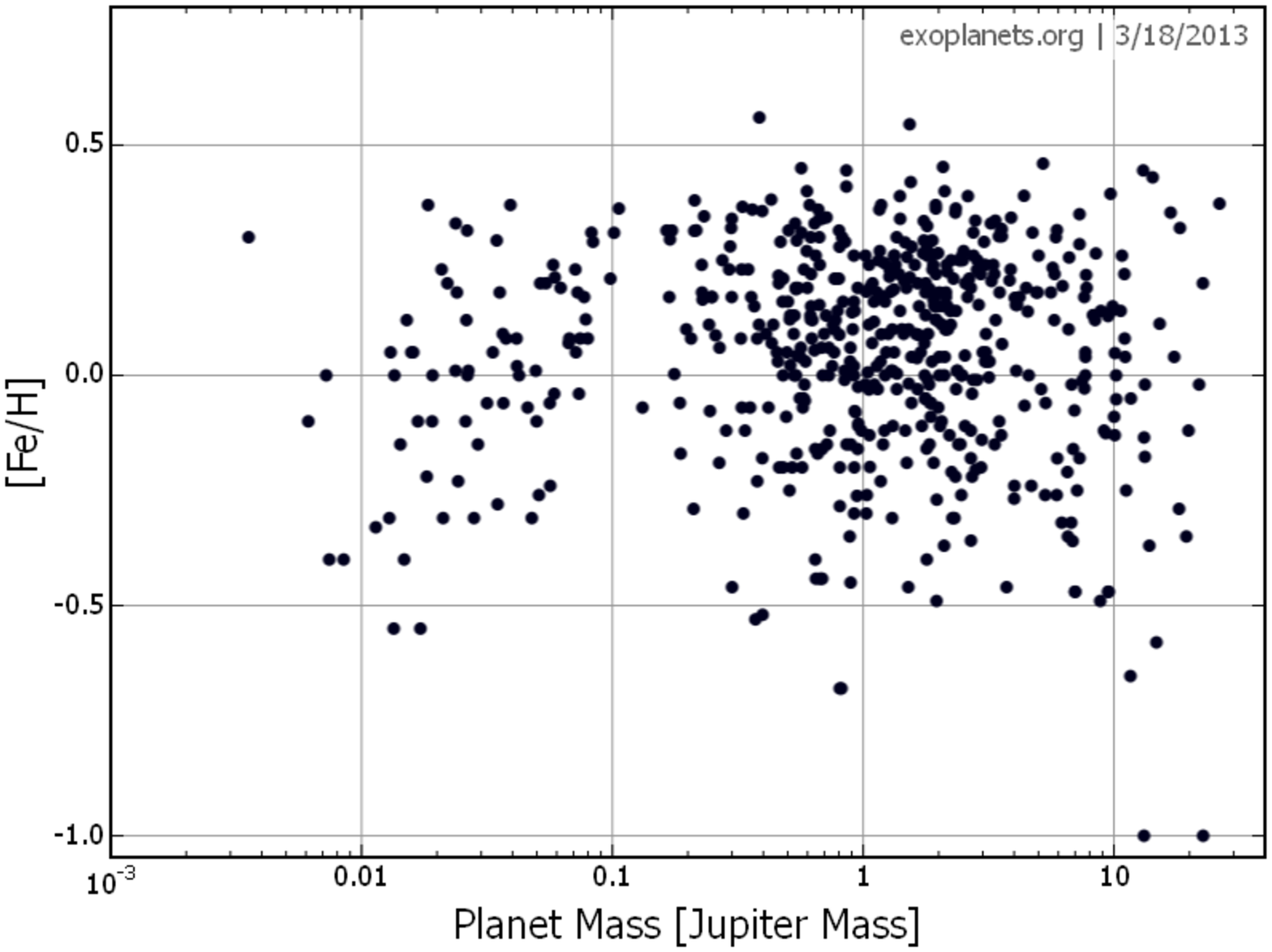} 
\includegraphics[scale=0.36]{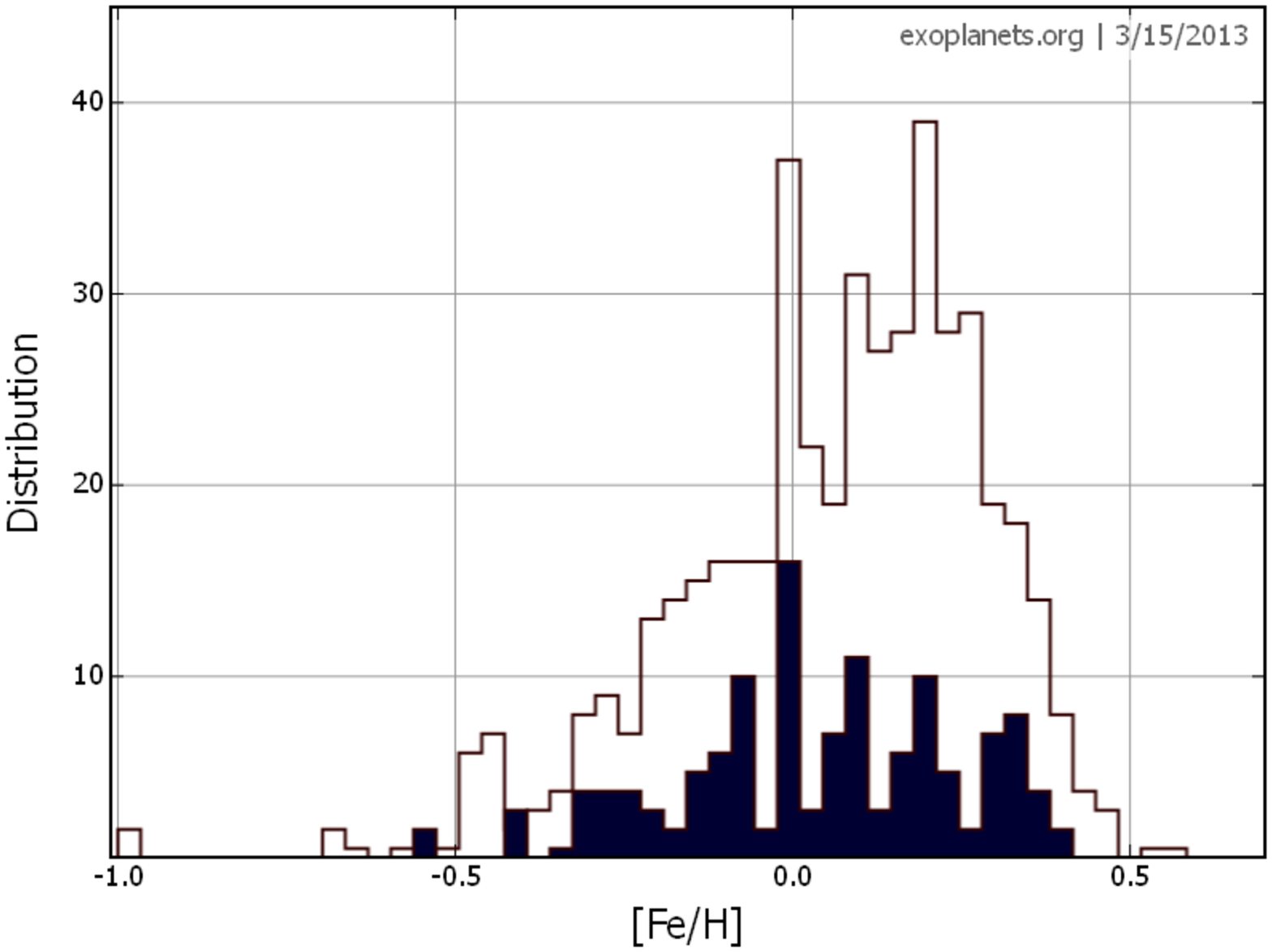} 
\caption{The left plot shows the distribution of the exoplanets masses vs. host metallicity (603 total). 
Two distributions are clearly seen: small-mass planets ($M< 0.3\, M_{\rm J}$) 
and giants ($M \geq 0.3\, M_{\rm J}$). The histograms for these distributions are plotted on the right.
Here, filled bars are the small-mass planets and open bars are the giants. These distributions are 
different at 92\% confidence level (see text). The histogram for giant planets shows a hint of a bimodal 
distribution with the peaks at [Fe/H]$\simeq 0$ and [Fe/H]$\simeq 0.2$. This figure was made using 
the Exoplanet Orbit Database. Notice small discrepancies between the two databases we used; Exoplanet 
Orbit Database does not show the metal-poor planets between -1.5 and -2.09. For KS test,
we used the data from Table~\ref{table:poor}.}
\label{fig:giants}
\end{center}
\end{figure}

In addition, the metallicity effect seems to be limited to giant planets. 
The recent population synthesis study by Mordasini et al. (2012) concluded that the metallicity 
acts as a threshold for formation of massive planets, i.e. in other words high metallicity allows 
the formation of a larger number of high-mass planets, adding to the natural observational bias 
since giant planets are easier to detect. The current ratio\footnote{according to Extrasolar Planets 
Encyclopaedia data of March 2013} of giant ($M\geq 0.3~M_J$) to small-mass ($M<0.3M_J$) 
planets is $\sim 3.1$, $M_J$ is Jupiter mass (Fig.~\ref{fig:giants}). In the lower mass 
domain --- Neptune masses --- this effect weakens (e.g., Mordasini et al. 2012). 
The host stars of Neptune-mass planets with precisely measured radial velocities 
(Cumming et al. 2008) show apparently a flat metallicity distribution (Udry et al. 
2006; Sousa et al. 2011): a Gaussian peak at $\rm [Fe/H]=-0.5$ transforms to a wide 
plateau in the range $\rm [Fe/H]=-0.9$ to $-1.5$. No metallicity trend is predicted 
for even lower-mass planets (Mordasini et al. 2012); moreover, recent results from 
Kepler mission (Buchhave et al. 2012) demonstrated that formation of small terrestrial-size 
planets does not necessarily require a metal-enriched environment.

On the left panel of Fig.~\ref{fig:giants} we show the distribution of exoplanets masses 
vs host metallicity, where two distinct mass populations are clearly seen. The host metallicity
distributions for these two populations are shown on the right panel of Fig.~\ref{fig:giants} 
(131 small planets are represented by filled bars and 472 giants by open bars). The distribution 
of giants has a median $\sim 0.1$ shifted to higher than solar values, suggesting that higher 
metallicities in the host stars is conducive for formation of giant planets -- the observed 
``metallicity effect''. Kolmogorov-Smirnov (KS) test performed to compare these two distributions 
shows that they are different at a $P$ value of $\sim 0.92$ (92\% confidence). Moreover, 
the distribution of small planets has a median of 0.035, which is closer to the solar value, 
and is visually more symmetric (has less skewness towards higher metallicities). 
The ``metallicity'' effect therefore seems to be absent for the smaller mass planets. 
In other words, smaller mass planets do not require for their formation the high-metallicity 
hosts, which is also concordant with the results from Kepler data (Buchhave et al. 2012).

\subsection{Planets around metal-poor hosts}

That planets can form from matter with lower than solar metallicity is evident from
the fact that about 50\% of the giant planet-host stars have sub-solar metallicities 
(Setiawan et al. 2012). With the launch of the surveys specifically 
directed at the metal-poor populations (Santos et al. 2007, Sozzetti et al. 2009, Setiawan et al. 2010, 
Klement et al. 2011), 
planets now are being found around hosts with fairly low metallicities, $\rm [Fe/H] \lesssim -0.4$.
 
It is interesting that when the first population statistics on exoplanets was published
in 2007 (with very few low-metal hosts known), the relative percentage of metal-poor stars 
($\rm [Fe/H] \lesssim -0.6$ with planets in solar neighbourhood ($\sim 4\%$) was 
four times higher than the relative percentage of Pop.~II stars ($\sim 1\%$) (as inferred from 
the Fig.~9 in Udry \& Santos, 2007; a similar number can be drawn from more recent study 
by Casagrande et al. 2011, Fig.~15). According to the latest data\footnote{Exoplanet 
Orbit Database, March 2013} this percentage now is $\sim 1.5\%$ (8 out of 535 stars 
with measured metallicity within $\sim$600 pc of the Sun have $\rm [Fe/H] \lesssim -0.6$, see 
compilation in Table~\ref{table:poor}). With more 
discoveries, the relative number of metal-poor hosts may asymptotically approach the 
relative number of Pop.~II stars in the solar neighbourhood. Thus, the question of 
principal importance arises of how to collect sufficient amount of metals to form planets, is it 
possible at all, and what could be the scenarios?

\begin{table}[ht!]
\begin{center}
\caption{Planets around stars with $\rm [Fe/H] \lesssim -0.6$}
\begin{tabular}{lllcclc}
\hline
 Name			&[Fe/H]		       &  Planet(s) mass &Star(s) mass&Sp. &Age & Ref. \\
                          &                        & (jupiter)             & (sun)         &Type       & (Gyr)   &      \\
\hline
HIP-13044 b        &$-2.09\pm 0.26$ 	& b=1.25                & 0.86      & F2        & $>9$  &    1         \\
HIP-11952 b,c      & $-1.95\pm  0.09$&  b=0.78, c= 2.93   &0.83       & F2V      &$12.8\pm 2.6$ &  2 \\
PSR-B1620-26 b   &$-1.5$		  	&b=2.5                   &1.35/0.6 &PSR/WD & 12.7  &       3        \\
BD20-2457 b,c    &$-1.0\pm 0.07$	&b=12.47, c=21.42   &2.8        & F9II    & pop.~II	&     4       \\
HD-114762 b       &$-0.71\pm 0.08$   &b=11.64                & 0.84      & F9V    & prob. old &    5        \\
HD-47536 b,c      &$-0.68$		       &b=5.0, c=7.0         & 0.94      & K1III   & $9.33\pm 1.88$  & 6 \\
HD-155358 b,c    &$  -0.68\pm 0.07$  &b=0.85, c=0.82      & 0.89      & G0      & 11.9   &     7          \\
AB-Pic b             &$-0.64$			&b=13.5                  & -           & K2V     & 0.03  &      8         \\
HD 13189 b         &$-0.58\pm 0.04$   &b=14.8                  & 4.5        & K2II     & Gyr  &      9         \\
\hline
\end{tabular}
\label{table:poor}
\end{center}
\vskip -0.1in
{\it Note:} Total: 9 hosts, 13 confirmed planets, all giants.\\
{\it References to discovery papers:}\\
{\bf 1}. Setiawan et al. (2010); {\bf 2}. Setiawan et al. (2012); {\bf 3}. Thorsett et al. (1993); 
{\bf 4}. Niedzielski et al. (2009); {\bf 5}. Latham et al. (1989); {\bf 6}. Setiawan et al. (2003);
{\bf 7}. Cochran et al. (2007); {\bf 8}. Chauvin et al. (2005); {\bf 9}. Hatzes et al. (2005).
\end{table}

\subsection{Planets in the early Universe}

In this note we argue that the conditions for selective assembling of metals in the dust form 
could have existed even in the very early Universe immediately after the formation of first
stars, i.e. at the epoch when the first stellar light illuminated the Universe at redshifts 
$z\gtrsim 10$ (see, e.g., Scannapieco et al 2006). It is worth mentioning in this connection 
that very recently Wickramasinghe et al. (2012) presented arguments that primordial 
free-floating planets of solid hydrogen may account the whole ``missing baryons'' in the Universe. 
When recalculated for the Milky Way, the number of such primordial planets shall be as 
numerous as $\sim 10^{14}$. In this work we however address rather the question 
of whether planets can form in low-metallicity environment. Moreover, we do not 
consider possible scenarios of planet formation in such conditions, but focus primarily on even 
simpler issue: what mechanisms can provide accumulation of sufficient amount of metals 
(mostly in solid phase) to make the subsequent formation of planets possible within scenarios 
similar to those acting in the local Universe. 

In Section~2 we bring evidence that in the very beginning of 
stellar nucleosynthesis in the Universe, one might find enriched regions with the metallicities close to 
solar, or even higher, where planets can form under conditions analogous to the ones in 
local Universe, provided that dust-to-metal ratio is of $\sim 0.3$ (as is normally assumed,
see \citet{malino04} and \citet{f11}). In Section~3 we describe a possibility of enhancement of 
dust abundance in metal-deficient gas either due to the centrifugal force in peripheral regions of 
circumstellar protoplanetary accretion disks, when the specific angular momentum is sufficiently high, 
or due to the action of the radiation pressure from the host protostar in the conditions of small 
specific angular momentum. Section~4 summarizes our results.

We have used for this discussion the observational data from the Extrasolar Planets Encyclopaedia 
({\tt http://www.exoplanet.eu}) maintained by J. Schneider, from the Exoplanet Orbit Database ({\tt http://www.exoplanets.org}) \citep{exoplanet.org} and from NASA Exoplanet Archive
({\tt http://exoplanetarchive.ipac.caltech.edu}). For the planet hosts with no metallicity data from 
the above-mentioned websites, we have used the values from Maldonado et al. (2012).

\section{Metallicity in the early Universe} 
\label{sec:2}

It became clear from the very first detections of metals at high redshifts $z\simeq 6$ (Songaila 2001) 
that there are extreme variations of metallicity in the Universe. Further observations confirmed this conclusion. 
Schaye et al. (2003) demonstrated that the abundance of carbon in the intergalactic medium can 
vary up to $2-3$ orders of magnitude, particularly at lower redshifts ($z=1.5-2.5$), while Simcoe et al. 
(2006) found many clumps and large clouds with metallicity varying between $\rm [X/H]=-3$ and $0.6$
toward quasar HS1700+6416 ($z_{\rm abs}=2.73$), here X denotes mostly the abundances of carbon, 
silicon, magnesium ions. Schaye et al. (2007) inferred a wide range (from $\rm [X/H]=-1$ to 0.5) of 
metallicities in the absorption systems of several quasars at redshifts of 
$z_{\rm abs}=2$ to $z_{\rm abs}=3.2$. They concluded that the metallicity tends to be higher 
for smaller sizes of the absorbing gas clouds, as is predicted by  the numerical simulations of 
metal-mixing by Dedikov \& Shchekinov (2004). 

From the theoretical point of view three basic models explaining the inhomogeneous metal 
distribution in the Universe have been described: patchy (unpercolated) distribution of 
metal-enriched bubbles around active (starburst) galaxies (Nath \& Trentham 1997; Gnedin 1998; 
Ferrara et al. 2000); dependence of the metallicity of enriched bubbles on the host galaxy 
(Nath \& Trentham 1997; Gnedin 1998; Ferrara et al. 2000; Shchekinov 2002), and inefficient 
(incomplete) mixing (Dedikov \& Shchekinov 2004).  

More evidence for highly inhomogeneous metal distribution at high redshifts comes from observations 
of Gamma-Ray Bursts (GRB) and particularly their optical afterglows: optical absorption spectra of the 
so-called GRB damped Ly-$\alpha$ systems -- as opposed to QSO damped Ly-$\alpha$ systems -- 
and the stellar light from GRB hosts show a high spread of metal abundance of about two orders of magnitude, 
with a weak redshift evolution and a relatively high mean $\rm [X/H]=-0.7$ (Savaglio 2008; Savaglio et al. 2009), 
suggesting that the metal distribution is highly inhomogeneous in the host galaxies. Numerical simulations 
of mixing of metals in the ISM of the Milky Way Galaxy (de Avillez \& Mac Low 2002) 
have shown that it can take as long as 120 Myr on length scales 
of 25 to 500 pc. If this time-scale is also applicable to GRB hosts, then homogenization of metals would 
take a considerable fraction of the Hubble time at $z=8$. The high spread of metallicity in the GRB hosts then 
simply reflects the inhomogeneity of their ISM. Under such circumstances, a wide range of physical 
conditions favourable for the formation of planets around hosts with solar or near-solar metallicity 
can be found in the early Universe.

\section{Low-metallicity environment} 
\label{sec:3}

Although we have shown that high-metallicity regions favourable for forming planets 
may be common in the early Universe, the independence or even possible anti-correlation
of the frequency of small planets with metallicity (Mordasini et al. 2012; Buchhave et al. 2012),
together with the existence of old metal-poor stars hosting planets, suggests that 
metallicity may not be the governing parameter for the formation of planets. 
It seems therefore worthwhile to consider possible mechanisms which can promote formation of 
planets in a metal-poor gas.

Core accretion scenario is recognized to account for most of detected planets. However, 
recent discoveries of planets which lie far from the possible parameter space for core accretion, 
for instance, planetary mass-metallicity relation, indicate that there may be several formation 
scenarios that lead to systems very different from the usual (see e.g., Mordasini et al. 2012). 
Such planets (for example, HD~155358, BD+20~2457 or HD~47536, see Table~\ref{table:poor}) 
are difficult to account for by the core accretion. Planets found around stars with low 
metallicities or on very wide orbits 
pose a challenge to the core-accretion scenario, and a gravitational instability model, where 
planetary frequency does not strongly depend on metallicity (Boss 2002), was proposed to 
better explain their formation (Dodson-Robinson et al. 2009). It is however possible that stars with 
different metallicities form their planets by different mechanisms. Such an assumption is 
consistent with the findings that both giant and Neptune-mass planets discovered so far have 
rather flat metallicity distribution 
(Santos et al. 2007; Udry \& Santos 2007; Sousa et al. 2008, 2011; Mayor et al. 2011). 
If the dominant planet formation mechanism were not changing with metallicity, we would 
rather see a monotonically increasing detection rate with [Fe/H] for all metallicity range 
(Dodson-Robinson 2012).  

Using order of magnitude estimate, we argue that, in the process of star formation from 
metal-underabundant material, immediately after the primeval enrichment in the early Universe, 
selective assembling of dust around protostars can operate and enhance the local dust 
abundance to a sufficient level so that physical mechanisms (such as dust coagulation) 
leading to planet formation can start operating. At this stage, we believe that such an order 
of magnitude calculation is more practical than a sophisticated numerical approach involving 
the entire evolutionary path from dust coagulation through planetesimals to planet formation --- 
a scenario which is still not fully understood even for our own Solar System.

\subsection{Gravitational collapse versus gas metallicity}
\label{sec:collapse}

Two key ingredients controlling formation of planets are the presence of dust in a star-forming medium 
and the amount of rotation. It was recently recognized (see, for ex. Toddini \& Ferrara 2001; Nozawa 
et al. 2003 and Maiolino et al. 2004) that dust particles are injected along with metals right from the 
very initial metal-pollution episodes. We therefore assume, following \cite{f11}, the mass abundance 
of dust to be ${\cal D}\simeq 0.3Z$, where $Z=Z_\sun \, \mathfrak{z}$ is the absolute mass abundance of metals.

It is obvious that in early galaxies with metal-poor gas, the low amount of metals and dust restricts 
the formation of rocky planets. The maximum number of Earth-size planets that can form in 
a protostellar cloud of mass $M$ is ${\cal N}\sim {\cal D}M/M_E$ where $M_E$ is the Earth mass. 
Defining $m=M/\msun$, we obtain 
${\cal N} \sim 3\times 10^3 m \mathfrak{z}$ per cloud. For the protostellar cloud of mass $m=10$ and 
even at $\mathfrak{z}\sim 10^{-2}$ and efficiency of planet formation of 1\%, we obtain ${\cal N}\sim 3$. If 
we assume the star formation efficiency at 10\% (i.e. each cloud of $m=10$ produces one star), it would mean 
that every star may have 3 planets. Note that when applied to the cloud with solar metallicity, this estimate 
gives up to 300 planets per star. Although it was recently recognized that multiplanetary systems are rather a 
rule than an exception, the most populous exoplanet systems yet detected, HD 10180, has 7 confirmed planets 
(Lovis et al. 2011). It is likely that the population of planets in a given system is usually less than 10 
(more likely 3)
because of the instability of planetary orbits on time scales greater than 1 Myr (Juric 2006; Veras et al. 2011).
The detection of multi-planetary systems can, however, be impeded by the high inclination of the
planetary orbits to the line of sight (Tremaine \& Dong, 2012) and/or by the weak and tangled 
wobbling of the host star produced by distant planets.

It is assumed after Hoyle (1953), that during gravitational contraction a cloud radius $R$ can be subjected to 
a successive (hierarchical) fragmentation, if radiative cooling is sufficiently efficient to keep temperature growing 
slower than $T\propto n^{1/3}$, such that the Jean's mass $M_J\simeq 30 T^{3/2}n^{-1/2}~M_\odot$ decreases 
with gas density $n$. More recent numerical simulations showed that successive fragmentation continues while 
gravitational contraction is isothermal, i.e. $\gamma=d\ln{p}/d\ln{\rho} \leq 1$ \citep{l03}. In general, 
since optical depth grows as $\tau\propto R^{-2}$, a contracting cloud becomes eventually opaque, even though 
at initial stages it is always optically thin. This moment finalizes the isothermal regime and determines the minimum 
stellar mass. 

A more recent development of this approach applied to the problem of Pop.~II.5 stars (stars formed right
after Pop.~III) was developed by Omukai et al. (2005)  and \citet{s06,so10,sch12}. They considered a parcel 
of gas density $\rho(t)$ belonging to a contracting fragment whose temperature is determined by heating from 
gravitational compression, and cooling is determined by radiation liberated in inelastic collisions,
\begin{equation}
\frac{1}{\gamma-1}\fr{k}{\mu m_{\rm H}}\fr{dT}{dt}=-p\fr{d}{dt}\fr{1}{\rho}-L\,,
\label{tem}
\end{equation}
where $m_{\rm H}$ is the hydrogen mass, $\mu$ is the mean molecular weight and $L$ is the cooling rate. 
Eq.~(\ref{tem}) determines the evolutionary track $T(\rho)$. For a given model of dust particles 
and their mass fraction, and when all possible    
cooling mechanisms, such as cooling in atomic (CI, CII, and OI) and molecular (H$_2$, HD, OH, CO) 
emission lines, cooling in continuum due to thermal emission by dust particles and others are included, 
the minimum mass of protostellar fragments is $\sim 0.3~M_\odot$ for metallicity $\rm [X/H] =-7$, and 
$\sim 1~M_\odot$ for $\rm [X/H]=0$ (Omukai et al. 2005). Therefore, within these assumptions, 
the masses of protostellar condensations are rather insensitive to gas metallicity from which they form, 
but can be sensitive to a dust model and to the depletion factor of refractory elements on dust surface 
\citep{sch12}. For simplicity, we restrict ourself with the models described by Omukai et al. (2005)
and Schneider et al. (2006). 

The cloud density at which fragmentation stops is, however, rather sensitive to metallicity and can be roughly 
approximated as $n\sim 10^5 \mathfrak{z}^{-1.5}-10^2 \mathfrak{z}^{-2}$ cm$^{-3}$, 
depending on the mass of a progenitor of the polluting supernova (SN), $195\,M_{\sun}$ and $22\,M_{\sun}$, 
respectively (Schneider et al. 2006). This can mean that less-enriched protostellar clouds and, as a consequence, 
the circumstellar (protoplanetary) disks when established, must rotate faster by factor of $n^{2/3}$ if they had 
equal specific angular momentum $j_0$ on average. The specific angular momentum of star-forming molecular 
clouds in the Milky 
Way is $j_0\sim 30$ pc km/sec, nearly an order of magnitude lower than the specific angular momentum in the 
interstellar medium \citep{b99}, which we will assume here as a reference value $j_0=10^{25}\Lambda$ 
cm$^2$/sec, with $\Lambda$ as a free parameter. 

\subsection{Centrifugal assembling of accreting dust}
\label{sec:4}

Planets seem to form during subsequent evolution of such a protostellar cloud, after formation of embryo disk 
through coagulation, settling of dust onto plane and followed fragmentation. The Toomre parameter for 
the stability of disk (Binney \& Tremaine 1987),
\begin{equation}
Q=\fr{\Omega c_s}{\pi G\Sigma}\,,
\label{eq:toomre}
\end{equation}
seems to be relatively insensitive to metallicity as both angular velocity $\Omega$ and surface density $\Sigma$ 
scale with the radius as $R^{-2}$. $Q$ is sensitive to the free parameter $\Lambda$, 
which can vary in a wide range. The sound speed $c_s$ at the time when fragmentation stops and accretion disk 
is to be formed depends on metallicity, $c_s\propto \mathfrak{z}^{-1/4}$, and can vary from 0.3 to 3 km/sec for 
$\mathfrak{z}=0.1 - 10^{-6}$ (as inferred from \citet{o05} and \citet{s06}). Later on, when the circumstellar 
accretion motion settles on to the disk, the sound speed relaxes to the value determined by the heating from central 
star (protostar) and by the viscosity in the disk. From this point of view, 
and particularly accounting for the fact that $\L$ varies in a wide range and might be low, one 
can conclude that the conditions for gravitational instability in circumstellar disks today and in 
the early galaxies are similar.  

At the initial stages when the disk has only just formed, dust particles move tightly coupled to gas. 
This is a fast rotation with velocity $v_\phi=\Omega R$ and a slow inward accretion with velocity $v_r\sim \alpha 
v_d^2/\Omega R$, $\alpha\leq 1$, where $v_d$ is the gas velocity dispersion (see Pringle 1981). Assuming the 
turbulence in the disk to be subsonic, let $v_d \leq c_s \propto \mathfrak{z}^{-1/4}$. Dust decouples 
from gas when the collisional drag rate, 
\be
\nu_d \sim \fr{\pi a^2 m_{\rm H} v_d n}{m_d} \propto a^{-1}\,,
\label{eq:dragrate}
\ee
becomes smaller that the rotation rate $\Omega$ due to the growth of dust particles; $m_d$ and $a$ are the grain 
mass radius, respectively. The total rate of the dust growth $\nu_g$ through coagulation 
and freezing of heavy elements on dust surface can be estimated as  
\begin{equation}
\nu_g(a) = \fr{1}{a}\dder{a}{t} \sim \fr{s m_H v_d n Z}{ a \bar\rho}\,,
\label{eq:rate}
\end{equation}
where $\bar\rho_g\sim 3$ g/cm$^3$ is the density of a dust grain and $s\simlt 1$ is the mean 
sticking coefficient in dust-dust and atom-dust collisions; tight collisional coupling between 
dust and gas is explicitly assumed. After time $t=2\pi N/\Omega$, corresponding to $N$ 
rotational periods of the disk, the grain radius 
can grow up to 
\be
a \sim \fr{sm_Hnv_d Z}{\bar{\rho}}\fr{2\pi N}{\Omega}\,,
\label{eq:rate2}
\ee
where we explicitly assumed that all dust particles are of equal radius. With the growth of dust grains,
the collisional drag rate $\nu_d$ decreases as 
\be
\nu_d\sim \fr{\Omega}{2\pi s Z N}\,,
\ee
so that after $N \gtrsim (s Z)^{-1}$, $\nu_d/\Omega \ll 1$, dust particles decouple from the gas motion.
Though they continue the frozen rotation, the inward radial motion slows down dramatically, 
$v_{dr}/v_{r}\sim O(v_r^2/v_\phi^2)$, $v_{dr}$ is the radial dust velocity, and dust piles up in a ring. This happens when the dust 
particles have moved inwards by 
\be 
\Delta R\sim \fr{2\pi Nv_r}{\Omega}\sim 2\times 10^{17} \fr{m}{\Lambda^2 n s Z}, 
\ee
where $m$ is the mass of a protostellar cloud in solar units (Sec.~3.1). The gas inflow continues, but its 
faster decoupling ($\Delta R/R$ is small) implies that it is left behind at larger radii. 
When the density at last fragmentation 
episode $n\propto \mathfrak{z}^{-b}$ ($b=1.5-2$, Schneider et al. 2006) is substituted, 
it becomes readily seen that after decoupling, dust particles in metal-deficient disks accumulate 
mostly in outer regions of the disk as $\Delta R\propto \mathfrak{z}^{b-1}$, i.e. the lower 
the metallicity of the disk the farther in the disk dust accumulates,
everything else (e.g. $\Lambda$, $m$) being equal. For the abundance pattern and dust yield corresponding 
to a $M=22\,M_{\sun}$ progenitor, we numerically obtain 
\be\label{dec}
\fr{\Delta R}{R}\sim \fr{0.1m^{2/3}\mathfrak{z}^{1/3}}{\Lambda^{2} s}\,,
\ee
which shows that in metal-deficient circumstellar disks dust accumulates in external regions, while in the cases 
with $\mathfrak{z}\sim 1$ dust occupies nearly the entire disk. Asymptotically, the enhancement factor in the 
region of accumulation is approximately $\eta\sim \Delta R/\delta R$; here $\delta R$ is the thickness of the 
ring where dust accumulates, estimated as the distance travelled by dust particles in the radial direction 
before being decoupled, $\delta R\sim ~3\times 2\pi v_r/\Omega$, a factor of 3 is assumed for concreteness. 
The enhancement factor is then readily estimated as $\eta \sim (3sZ)^{-1}$, resulting in the enhanced 
metallicity in the ring $\mathfrak{z}_e\sim\eta\mathfrak{z}\sim (3sZ_{\sun})^{-1}\simgt 1$. It has 
to be stressed that formation of ring-like structures in an accretion flow due to the centrifugal force 
occurs only in metal-poor environment with $\Delta R/R\propto\mathfrak{z}^{1/3}$, under the assumptions specified 
above such as SNe dust production scenario, formation of stars via hierarchical fragmentation, etc. Observations of 
dust rings in accretion debris disks in the local Universe (see, e.g. Fitzgerald et al., 2007) reveal 
phenomena of different nature, most likely connected with the influence of radiation of the central star, in particular,  
dust-mantle sublimation. In our case, the sublimation of icy dust mantles can also take place at distances smaller 
than 10 AU; however, centrifugally-supported dust rings are to form, most likely, at much higher distances of 
thousands of AU.

After decoupling, the gas is at a higher temperature and resists gravitational collapse,
while the dust has smaller velocity dispersion and is thus able to form gravitationally-bound objects. 
The Toomre parameter of dust $Q_d$ within the ring decreases inversely proportional to 
dust content in it and, because the dust velocity dispersion $\sigma_d$ ($\sigma_d \ll R\Omega$)
decreases as $\sigma_d\propto a^{-3}$,
\be
Q_d=\fr{\Omega \sigma_d}{\pi G\Sigma_d}\sim 0.1{\Lambda v_{d}\over m}\left(\fr{a_0}{a}\right)^3\,,
\ee
where $a_0$ is the initial radius of the grain and we have assumed that before decoupling dust and 
gas velocity dispersions are equal, $\sigma_d=v_d$, $v_d$ is in cm~s$^{-1}$. It can be 
readily shown that during the growth stage, the grain radius $a$ increases considerably by factor 
$a/a_0\sim 50\mathfrak{z}^{-1/3}m^{2/3}$. This means that $Q_d$ gradually falls below unity.  
As a result, dust particles in the ring start assembling due to self-gravity, with characteristic free-fall
time 
\be
t_{ff,d}=\sqrt{\fr{3}{32\pi G\rho_{d,e}}}\,,
\label{eq:ffdust}
\ee
where $\rho_{d,e}$ is the enhanced dust density in the ring. It can be expressed as 
$\rho_{d,e}= 0.3\rho Z_{\sun}
\, \mathfrak{z}_e$, where $0.3\, Z_{\sun}\,\mathfrak{z}_e$ is the mass fraction of dust in the ring
and $\rho\sim m_H n$, and $n$ is assumed to be close to the value at the final fragmentation 
(Omukai et al. 2005), i.e. $n\sim 10^2\,\mathfrak{z}^{-2}$. The free-fall time is then
\be
t_{ff,d} \sim \fr{10^{15}}{\sqrt{0.3 Z_{\sun}\,\mathfrak{z}_e n}} \sim 
\fr{10^{16}}{\sqrt{n\mathfrak{z}_e}} \simlt 10^{15}\mathfrak{z}\quad \mbox{sec}\,.
\label{eq:11}
\ee
It is readily seen that the assembling can take tens to hundreds of rotation periods, depending on 
the mass of the protostellar cloud, rotation free parameter $\Lambda$ and metallicity $\mathfrak{z}$. 

It is clear that the overall picture depends on the dust optical model, i.e., on the enrichment scenario. 
When, for example, the abundance pattern of a pair-instability SN with a $195\,M_\odot$ progenitor is applied, 
for which \cite{o05} and \cite{s06} predict the density at the final fragmentation step 
$n\sim 10^5 \mathfrak{z}^{-3/2}$, decoupling condition becomes insensitive to metallicity and dust 
occupies the whole disk. As a consequence, the dust distribution throughout 
the disk becomes very dilute, making its gravitational accumulation a very slow process. At such 
conditions, however, dust can be selectively assembled by the radiation from the central star 
through the ``mock gravity'' instability.

\subsection{Radiation-driven assembling of dust clumps}
\label{sec:5}

When either dust optical properties critically deviate from the model used in above estimates, or 
angular momentum is low $\Lambda\ll 1$, the described scenario of centrifugal dust accumulation does not 
operate anymore. In such conditions, a complementary mechanism, which selectively 
drives dust particles to form clumps, connected with radiation pressure from a central stellar or 
protostellar object, can come into play.
Let us discuss briefly this scenario assuming for the sake of simplicity  $\Lambda\ll 1$. When 
a dusty gas is illuminated by an external radiation field, a 
mock gravity instability can develop with the growth rate similar to the one of Jean's instability \citep{f71},
\be\label{gm}
\Gamma^2=f(\sigma)\Gamma_r^2-k^2c_{dg}^2\,.
\ee
Here $\Gamma_r=\kappa c(\rho_r/\rho)^{1/2}$, where $\kappa$ is the inverse length 
of photon extinction and $c$ is the light speed. Radiation density is 
$\rho_r=4\pi F/c^3$, where $F$ is the photon energy flux, $\rho=\rho_d+\rho_g$ is the total 
dust and gas mass density, and $k$ is the wavenumber, $\sigma=k/\kappa$, the normalized wavenumber, $f(\sigma)$  
is defined by
\be
f(\sigma)=\frac{1}{3}\sigma^2\Biggl[\frac{1-\beta}{\beta}+\frac{1}{1-\tan^{-1}\sigma/\sigma}\Biggr]\,,
\ee
where $\beta$ is the fraction of incident radiation absorbed by a dust grain. 
The sound speed in dust-gas mixture, $c_{dg}$, is essentially equal to the sound
speed in gas, $c_{dg}\simeq c_g$. It can be readily seen that for all wavenumbers of interest, 
$\sigma \gg 1$, 
in which case the asymptotics of $f(\sigma)\sim Q_a$, where $Q_a$ is the absorption efficiency \citep{f71}. 
In the long-wavelength limit, the wavenumbers are less than the critical wavenumber, $k<k_c=\Gamma_r/c_{dg}$, 
and perturbations in such dust-gas mixture become unstable against aperiodic growth of clumps under 
the action of radiation pressure.

The optical flux from a nearly-solar mass protostar at distance $r$ is 
\be\label{ps}
F \simeq \fr{pL_\odot}{4\pi r^2}\simeq 2\times 10^5~r_{_{\rm 10}}^{-2}\quad \mbox{erg cm$^{-2}$ sec$^{-1}$}\,,
\ee
where $r_{_{\rm 10}}$ is the distance from star in units of 10 AU and $p$ describes the deviation of the luminosity 
from the main-sequence (MS) solar luminosity before settling of a protostar on to the MS. In Eq.~(\ref{ps}) 
we explicitly assumed $p=10$ following conclusions by \citet{p93}. We can estimate the inverse extinction length as
\be
\kappa=\pi a^2{\cal D}Z_\odot\fr{\mathfrak{z}\mu m_{\rm H}n}{m_d},
\ee
where dust, as above, is assumed monodispersal. For typical parameters, we obtain the critical length 
\be
\lambda_c=\fr{2\pi}{k_c}=\fr{2\pi}{\kappa}\fr{c_g}{c}\left(\fr{\rho}{Q_a\rho_r}\right)^{1/2}=
\fr{8\pi\bar\rho a}{3{\cal D}Z_\odot\mathfrak{z}\rho}\left(\fr{\rho}{Q_a\rho_r}\right)^{1/2}\fr{c_g}{c},
\ee
here density of dust particles is assumed as $\bar\rho=3$ g cm$^{-3}$.

The corresponding critical mass is 
\be
M_c \sim \rho\lambda_c^3 \simeq \left(\fr{8\pi\bar\rho a}{3{\cal D}Z_\odot\mathfrak{z}}\fr{c_g}{c}\right)^3
\fr{1}{(Q_a \rho_r)^{3/2}}\fr{1}{\rho^{1/2}} \sim 
2\times 10^{31}\fr{r_{_{10}}^3T_{100}^{3/2}a_{0.1}^3}{n^{1/2}p^{3/2}\mathfrak{z}^3}\quad\mbox{g}\,.
\ee
Here $T_{100}=T/100~{\rm K}$ is the temperature in the circumstellar cloud and $a_{0.1}=a/0.1~\mu$m. 
If one assumes that the  
density in such a cloud, surrounding the central protostar, is close within a factor 
$\nu<1$ to the density at the time of last fragmentation (when the protostar has formed), 
the critical mass becomes
\be
M_c^{(1)}\sim 10^{26}\nu^{-1/2}\mathfrak{z}^{-9/4}r_{_{10}}^3\quad \mbox{g}\,, 
\ee
for dust composition related to a $\sim 195 M_{\sun}$ polluting SN, and 

\be
M_c^{(2)} \sim 3\times 10^{27}\nu^{-1/2}\mathfrak{z}^{-2}r_{_{10}}^3\quad\mbox{g}, 
\ee
for a $22 \,M_{\sun}$ SN dust composition, as inferred from \cite{s06}; here $T_{100}=0.1$ and $p=10$. 
In both cases, $M_c$ is close to the Jovian 
mass for $r_{10}=1$, when the metallicity range $\mathfrak{z}=0.1-0.01$ is assumed. Note that the 
corresponding mass of metals, $M_z=Z M_c=Z_\odot\mathfrak{z}M_c$, is either equal or even less than the 
mass of the Earth. It is interesting to note that the critical mass depends on the distance from the central 
protostar $M_c\propto r^3$, which is obviously caused by the fact that the instability is driven by the 
protostar radiation --- the farther from the protostar, the weaker is the instability, and the longer is the 
unstable wavelength: at $r=1$ AU the critical mass is already in the range of Earth masses. It is 
worthwhile to stress here that low specific angular momentum $\Lambda\ll 1$, that we have assumed in this section, 
does not mean that dust clumps would fall onto the central star and evaporate. Firstly, $\Lambda\ll 1$ implies 
only that the disks form at much later stages than described by Eq. (\ref{dec}), that is, dust does not 
decouple centrifugally. Secondly, it can be readily seen that the characteristic growth 
time of the mock gravity instability, $\tau_r\sim 10a_{0.1}r_{10}$ yr (for a $22~M_\odot$ dust composition of 
Schneider et al., 2006), is comparable (or shorter at radii $r_{10}>1$) to the free-fall time of clumps onto 
the star, $t_{ff}\sim 10r_{10}^{3/2}$ yr (Eq.~\ref{eq:11}). When, on the contrary, angular momentum is not 
necessarily $\Lambda\ll 1$, an additional positive energy -- epicyclic motion $\varkappa^2$ 
(Pringle \& King, 2007) -- enters the r.h.s. of Eq.~(\ref{gm}) along with the 
sound mode $k^2c_{dg}^2$, which increases critical lengths and masses of the fragments formed.

\section{Conclusions}
\label{sec:conclusions}

\begin{enumerate} 
\item{} Recently discovered planets around metal-deficient stars, such as HIP 13044 and HIP 11952,
indicate that planets can form in the early Universe. Free-floating planets 
observed in MOA experiment can also be old as they are located in the region of stellar 
Pop.~II (in this sense, they belong to Population II planets --- planets formed after the initial 
episodes of enrichment by Pop.~III stars, either {\it in situ} with Pop.~II hosts or {\it ex-situ} 
from direct gas collapse). At the same time, it cannot be excluded that a well-established 
increase of metallicity towards the Galactic center can contaminate this expectation. 

\item{} On the other hand, observations definitely show that metals in the Universe are mixed very 
inefficiently, resulting in a highly inhomogeneous distribution of metallicity in the Universe: 
it is clearly seen both in stellar populations of GRB hosts and in the IGM clouds of GRB 
damped Ly-$\alpha$ systems. As a consequence, even in a high redshift Universe (say, $z\gtrsim 6$), 
the conditions can exist for formation of planets with normal metallicities, either anchored to 
host stars or free-floating in the star-forming clouds.

\item{} Planets can also form in metal-poor environments: 

\begin{itemize}  

\item{} In the cases when the specific angular 
momentum in protostellar molecular clouds is comparable to the value of the Milky Way molecular clouds. 
In such case, they form preferentially in outer regions of circumstellar disks due to formation of dust 
rings, where dust particles are decoupled from gas and have negligible inward radial velocity. 

\item{} If, however, rotation of starforming clouds is small and circumstellar gas accretes 
quasi-spherically, the radiation 
from the central protostar can stimulate formation of clumps through the mock-gravity instability caused by 
radiation pressure on dust grains. The characteristic masses of such clumps for fiducial parameters are close to 
the Jovian mass. Such clumps can, in principle, give rise to the formation of (free-floating) planets in 
a metal-deficient environment of the early Universe.

\end{itemize}
\end{enumerate}
Therefore, the conditions favourable for planet formation --- sufficiently high dust-to-gas ratio ---
could have existed in the early Universe in both metal-rich and metal-poor environments, 
which can explain the existence of planets with an age of up to 13-14 Gyr in the local Universe. These conditions,
that we have envisaged here, are however only necessary though not sufficient, so that further observational and 
theoretical study are needed in order to firmly substantiate this conclusion.

\section{Acknowledgments}

YS is supported by the RFBR (project codes 09-200933, 11-02-97124p, 11-02-00263) and 
acknowledges the hospitality of IIA, Bangalore, when this work has been initiated. Authors 
thank Tarun Deep Saini for his valuable comments and continuous assistance, 
V. Zh. Adibekyan and A. V. Tutukov for comments. This research has made use of the NASA Exoplanet Archive, which is 
operated by the California Institute of Technology, under contract with the National Aeronautics 
and Space Administration under the Exoplanet Exploration Program, and of the Exoplanet Orbit Database
and the Exoplanet Data Explorer at {\tt exoplanets.org}. This research has made use of 
NASA's Astrophysics Data System Abstract Service.

\end{document}